# An automated image analysis framework for segmentation and division plane detection of single live *Staphylococcus aureus* cells which can operate at millisecond sampling time scales using bespoke Slimfield microscopy

Adam J. M. Wollman*[1], Helen Miller*[1], Simon Foster[2], Mark C. Leake[1,3]

*Authors contributed equally

1. Biological Physical Sciences Institute (BPSI), Departments of Physics and Biology, University of York, Heslington, York, YO10 5DD.
2. Krebs Institute, Department of Molecular Biology and Biotechnology, University of Sheffield, Sheffield, S10 2TN.

3. For correspondence: mark.leake@york.ac.uk

**Abstract**

*Staphylococcus aureus* is an important pathogen, giving rise to antimicrobial resistance in cell strains such as Methicillin Resistant *S. aureus* (MRSA). Here we report an image analysis framework for automated detection and image segmentation of cells in *S. aureus* cell clusters, and explicit identification of their cell division planes. We use a new combination of several existing analytical tools of image analysis to detect cellular and subcellular morphological features relevant to cell division from millisecond time scale sampled images of live pathogens at a detection precision of single molecules. We demonstrate this approach using a fluorescent reporter GFP fused to the protein EzrA that localises to a mid-cell plane during division and is involved in regulation of cell size and division. This image analysis framework presents a valuable platform from which to study candidate new antimicrobials which target the cell division machinery, but may also have more general application in detecting morphologically complex structures of fluorescently-labelled proteins present in clusters of other types of cells.

**Introduction**

The application of novel biophysics tools is generating important new insight into processes of infection (1–5); in particular, biophysical instrumentation in the form of bespoke light microscopy hardware, and of bespoke image analysis software tools to extract meaningful information from the images that are generated in an often low signal-to-noise ratio regime, is generating promising new understanding of the biological mechanisms which underlie the process of antimicrobial resistance in a range of different pathogens. An example of such a pathogen relevant to human disease is *Staphylococcus aureus,* a bacterium that reproduces through binary fission into cellular clusters. *S. aureus* is a normal member of human skin microflora (6,7) but causes serious infections on reaching underlying tissues. To study the process of *S. aureus* cell division, we combined several existing image analysis tools into a new framework which, for the first time, was applied to living *S. aureus* pathogens imaged at millisecond time scales to single molecule detection precision using the bespoke biophysical optical imaging technique of Slimfield microscopy. This analysis framework enabled us to detect the cell division plane of individual cells, their boundaries and other subcellular morphological features.

*S. aureus* infection of skin and lung may lead to advanced systemic bacterial infection, or bacteraemia, a fatal condition if the strain is resistant to antibiotics (8). Methicillin resistant *S. aureus* (MRSA) is resistant to beta-lactam antibiotics, such as those based on penicillins and cephalosporins, and most broad spectrum fluoroquinolones such as Ciprofloxacin (9,10). Antibiotic resistance is an enormous problem now in clinical treatment centres, especially so for surgical procedures involving joint replacement and secondary infections arising following chemotherapy. MRSA can be treated currently with the glycopeptide vancomycin, however strains have recently been identified with reduced susceptibility to vancomycin (11) and even complete resistance (12), so-called VRSA. Many traditional antibiotics operate through targeting of cell wall components in invading microbes. For example, beta-lactam antibiotics inhibit cell wall synthesis of peptidoglycan to disrupt the cell's ability to withstand high osmotic cellular pressure. They bind irreversibly to the active site of penicillin binding proteins, preventing them from building cross links in the cell wall (13,14). Resistance to beta-lactam antibiotics is however prevalent, having evolved into strains which have binding sites with significantly reduced binding affinity to the antibiotics, or have developed new forms of enzymatic degradation of the beta-lactam motif (15). Others such as the fluoroquinolones operate through targeting DNA replication. The process of cell division may present alternative molecular candidates for targeted disruption by newly developed antibiotics. However, cell division processes have been studied primarily in the model rod-shaped organisms *Bacillus subtilis* and *Escherichia coli*, which have less specific relevance to biomedically more harmful pathogens such as *S. aureus*.

Cell division in *S. aureus* is driven by a complex mix of several proteins, many essential, termed the divisome (16). The protein FtsZ forms a ring structure at a future division site at mid-cell, known as the 'Z ring'. The exact role of many of the proteins involved in cell division, and the extent of their essential nature or not in different organisms, are unknown. The protein EzrA (denoted so for 'Extra Z rings A') is crucial in *S. aureus* (17,18). In *B. subtilis* EzrA acts as an inhibitor, preventing the formation of multiple Z rings per cycle, and EzrA is also recruited to the mid-cell early in the cell division process (19). In *in vitro* assays, EzrA is observed to interact with the C terminus of FtsZ which prevents it from assembling the Z ring (20,21). The idea that an inhibitor of Z ring formation is recruited to the divisome is surprising, but in *S. aureus* EzrA was found to also regulate cell size (17,18), preventing cells from getting so large that the Z ring could not form correctly. This agrees with the finding that in *S. aureus* inhibition of cell division produces cells up to twice as large as normal (17,22).

The localisation of EzrA changes through the cell cycle. In *S. aureus*, EzrA is known to locate to the mid-cell early in the division process (17,23), during which period *S. aureus* becomes oblately ellipsoidal rather than truly spherical (24). EzrA can therefore be used as a useful marker for the cell division plane in the early stages of cell division, and thus its identification may provide a useful platform from which to study antimicrobial activity which affects cell division processes, or its absence in resistant strains.

Light microscopy has developed from its inception over 300 years ago into being an invaluable biophysical tool for studying complex biological processes in living cells (25). Similarly, automated analytical and computational tools of theoretical biophysics are proving useful in the interpretation of new forms of imaging data (26,27). In particular, the use of fluorescence microscopy, and associated analyses of the resultant images for studying complex processes, has added much to our understanding of complex molecular architectures inside living cells. For example, our previous work in this area includes studying cellular bioenergetics (28–32), protein transport (33–35), DNA replication and repair (36,37), cell movement and sensing (38–41), chromosome architecture (42–44) and developing new experimental and analytical imaging tools to probe general molecular machines in live cells at a single molecule precision (26,45–52).

Earlier attempts at monitoring cell division processes in *S. aureus* used labour-intensive manual image segmentation methods (53), highly prone to user subjectivity. Other attempts have utilised super-resolution images on fixed (i.e. dead) cells (24), limiting the study of dynamic biological processes. Although several standard methods already exist for the rough segmentation of bacterial cell clusters and the simultaneous detection of fluorescence from a labelled intracellular protein (54,55) these have never been applied to *S. aureus* cell clusters over a challenging millisecond time scale relevant to *in vivo* molecular mobility. Our image analysis framework uses fluorescence and brightfield image data as an input and interrogates these with automated image segmentation and watershedding algorithms to detect individual *S. aureus* cells, determine the location of their cell walls, and identify cell division planes in cells containing fluorescently labelled EzrA. Importantly, these techniques can be applied to imaging data from Slimfield microscopy (4,36,37,43,49) which enables tracking of single-molecule complexes over millisecond time scales which are comparable to diffusive molecular mobility inside living cells (27,56), as well as being compatible with advanced analytical methods which employ single cell copy number quantification through convolution modelling (57).

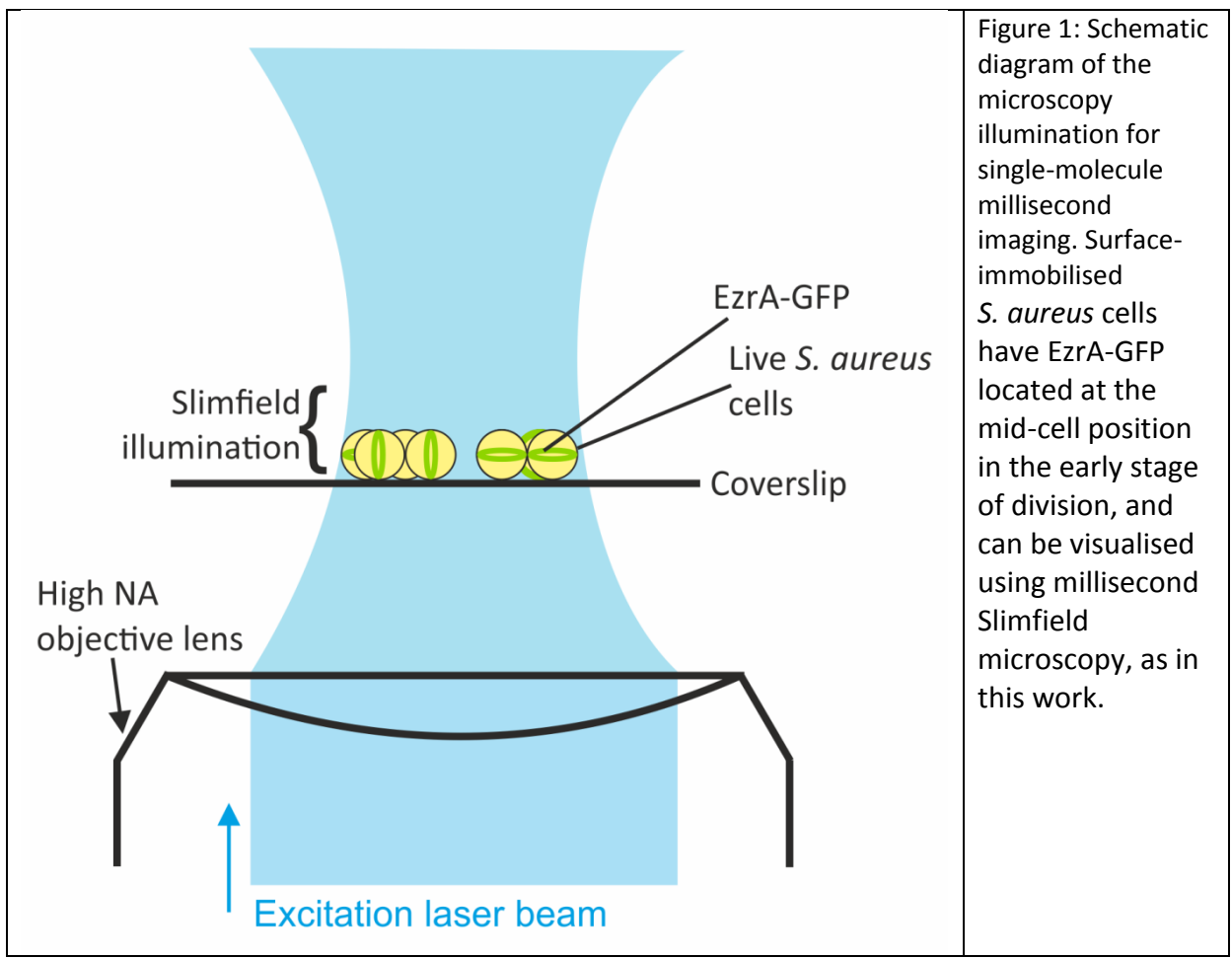

Figure 1: Schematic diagram of the microscopy illumination for single-molecule millisecond imaging. Surface-immobilised *S. aureus* cells have EzrA-GFP located at the mid-cell position in the early stage of division, and can be visualised using millisecond Slimfield microscopy, as in this work.

**Methods**

Cell strains

*S. aureus* cell strains SH1000 (the parental wild type strain used for native autofluorescence measurements) and SH1000 EzrA-GFP+ (EryR) were stored in glycerol frozen stocks at -80˚C. Cell cultures were grown as detailed previously for these strains (17) in rich media TSB (Tryptic Soy Broth; 17g Trypticase peptone, 3g Phytone peptone, 5g sodium chloride , 2.5g Dibasic potassium phosphate, 2.5g glucose, 1L deionised water, pH 7.3) at 37˚C.

Fluorescence microscopy

We used a bespoke single-molecule fluorescence microscope constructed around the chassis of a Nikon TE2000 inverted microscope using a 100x 1.49 NA oil immersion total internal reflection fluorescence (TIRF) objective lens (Nikon) and a *xyz* nano positioning stage (Nanodrive, Mad City Labs). Here we used fluorescence excitation from a 50 mW Obis 488nm laser to excite GFP fluorescence. A dual-pass GFP/mCherry dichroic mirror with 20 nm transmission windows centred on 488 nm and 561 nm was used below the objective lens turret. The beam was expanded to generate Slimfield excitation of 10 µm full width at half maximum in the sample plane of excitation intensity 1.5 $Wcm^{-2}$. Slimfield illumination operates by underfilling the back aperture of a high NA objective lens using a collimated laser beam (48). Underfilling results in a conflated confocal volume in the sample plane which can be adjusted by changing the upstream beam expansion optics to be marginally larger than a single cell, or cluster of cells, as required. In doing so the local laser excitation intensity is high enough to permit millisecond time scale image sampling of entire single live cells or small groups of cells with no requirement for slow scanning, at a detection precision to detect single fluorescent protein molecules while still producing an emission signal above the level of camera readout noise (4,35,36,48, 57). The Slimfield beam intensity profile was measured directly in a separate experiment by raster scanning in the focal plane while imaging a sample of 100 nm diameter fluorescent beads (Molecular Probes) immobilised to the coverslip surface. A high speed camera (Photometrix Evolve Delta) was used to image at 5 ms/frame (this is the time between consecutive frames, of which 0.6 ms is the 'dead' time in the 5 ms window during which the camera is unable to acquire data) with the magnification set at 80 nm per pixel. The microscope was controlled using Micro-Manager software (59).

Sample preparation and imaging

Microscope flow cells, or 'tunnel slides', for imaging were constructed from BK7 glass microscope slides (Fisher) and plasma-cleaned coverslips (Menzel Glaser) by laying two lines of double-sided tape (Scotch) approximately 5 mm apart on the slide and dropping a coverslip onto the tape and tapping down (avoiding the imaging region), to produce a watertight linear channel (60). The tunnel slide was coated in 0.01% poly-L-lysine to immobilise cells, and inverted for 5 minutes. This was then flushed through with 200 µl phosphate buffered saline (PBS). Following this, a tunnel volume of cells were flushed through and the slide was left inverted for 5 minutes to allow cells to attach to the coverslip. After 5 minutes any unattached cells were washed out with 200 µl PBS buffer prior to Slimfield imaging (fig. 1).

Imaging analysis framework – 1. Image segmentation of cells

Brightfield and fluorescence images were segmented initially by simple pixel thresholding. Here, we defined the background intensity from the modal (i.e. peak) value of the pixel intensity histogram as follows. The density of cells in a tunnel slide was optimized such that there were typically >80% more background pixels than foreground pixels (i.e. those associated with cells) (fig. 2a,b). This ensured a distinct modal peak in the pixel intensity histogram corresponding to the first fluorescence image in a kinetic series, which was associated with the local background value. We then used an automated thresholding method to find the pixel intensity values which were greater than the background peak value plus one full width half maximum (FWHM) of the background peak itself as a simple and automated initial method to discriminate the background from the foreground cell-containing regions, which contained one or more commonly more (up to ~10 cells) in a cluster in fluorescence images (fig. 2c). In the case of simple isolated single cells standard morphological transformations can in principle be used to fill holes in segmented regions and remove small objects and single pixels (57), however in general *S. aureus* cells appear in clusters, requiring further image segmentation.

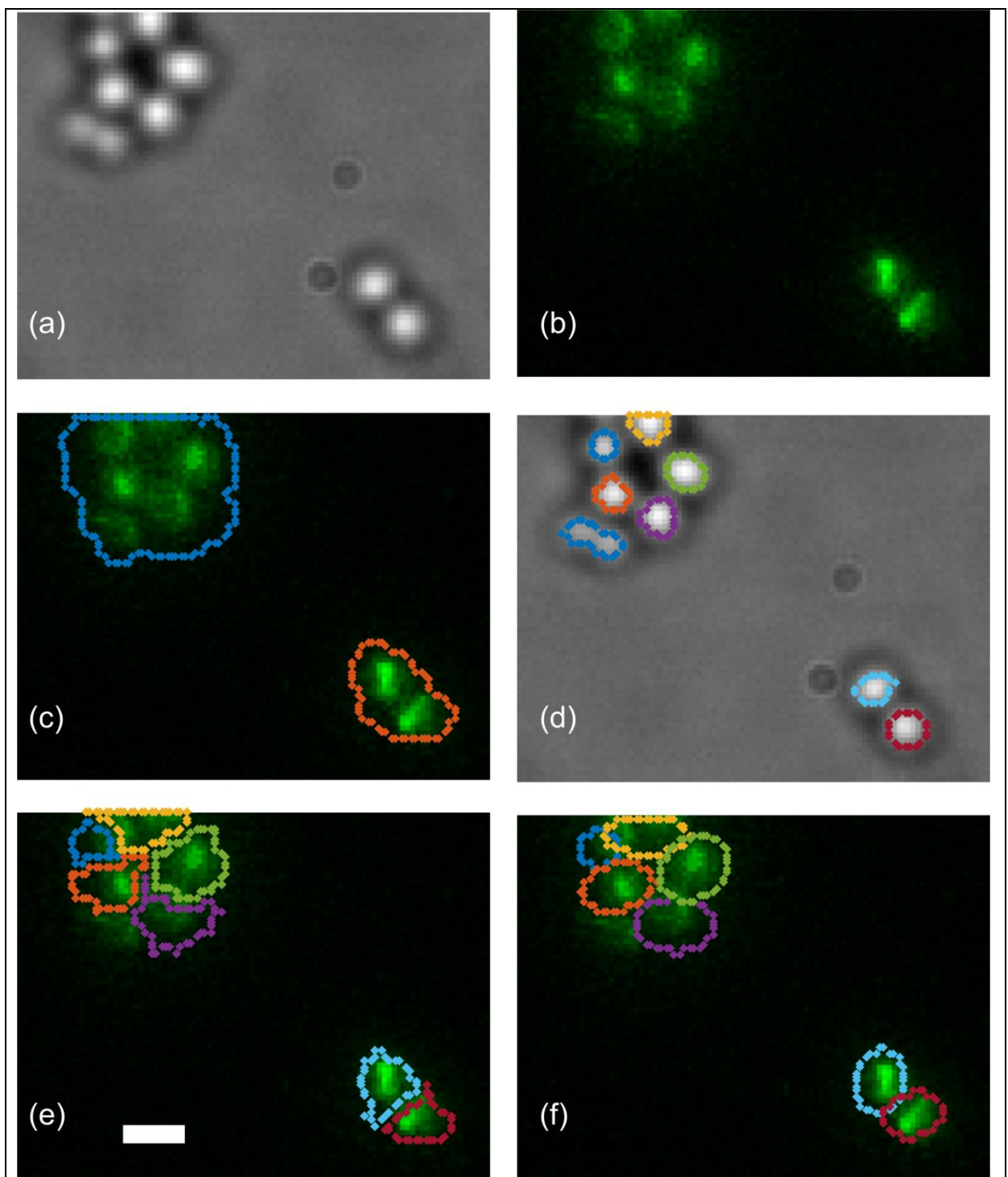

Figure 2: The cell segmentation algorithm (a) brightfield micrograph of staph cells, (b) false-colour fluorescence micrograph of EzrA-GFP, (c)segmented GFP image – cell containing regions are segmented, (d) segmented brightfield image providing individual cells and seeds for watershedding in (e), (f) watershedded cell masks used to generate ellipses.

Our simple brightfield images are slightly defocused by a few hundred nm compared to fluorescence images to provide greater image contrast, resulting in a dark ring appearance around the perimeter of cells which results in under segmentation artefacts if the simple pixel thresholding method is applied (fig. 2d). Segmenting the fluorescence image is advantageous as there is typically a low-level, uniform autofluorescence in cells due to the presence of natively fluorescent flavins and nucleotide derivatives (41,43,57) which gives a more accurate boundary between the cell image and the background (57). The disadvantage of this method is that close-packed cells, such as the clusters typical of *S. aureus*, are normally manifest as often contiguous regions of very similar pixel intensity since the cells are not separated by clear regions of background intensity (fig3c). For elongated objects standard methods to separate overlapping cells with cell-background boundaries exist (61); however, for cells with only cell-cell boundaries we found that using the brightfield segmentation outputs (fig3d) to define primary seeds in a watershedding algorithm allowed the separate cell boundaries to be determined accurately.

We developed software implementing these algorithms written in MATLAB (Mathworks) which automatically determined the segmentation pixel thresholds of the fluorescence image and brightfield image to determine images masks corresponding to the spatial extent of individual cell clusters (fig. 2c)), and individual cell seeds for watershedding (fig. 2d) respectively. The watershedding algorithm estimates which pixels are associated with each individual cell in a cell cluster (fig. 2e). This raw pixelated watershed segmentation output is then modelled as a 2D ellipse function, with fitting optimization generating estimates for the minor and major axes, centroid position and orientation of each segmented cell region (fig2f).

Watershedding algorithms derive their name from the concept of river catchment basins; the areas of land from which surface water will drain into that river. The ridges in the landscape form dividers (or watersheds) between adjacent catchment basins. Our primary image segmentation step uses the autofluorescence signal of the cells to first determine the outer boundaries of cell clusters, and we then apply a watershedding algorithm to find the borders of the individual cells within each cluster. By inverting the pixel values of the fluorescence image we generate a contour map such that the positions of the centres of cells correspond to valleys and the ridges between the valleys mark the cell boundaries (fig. 3).

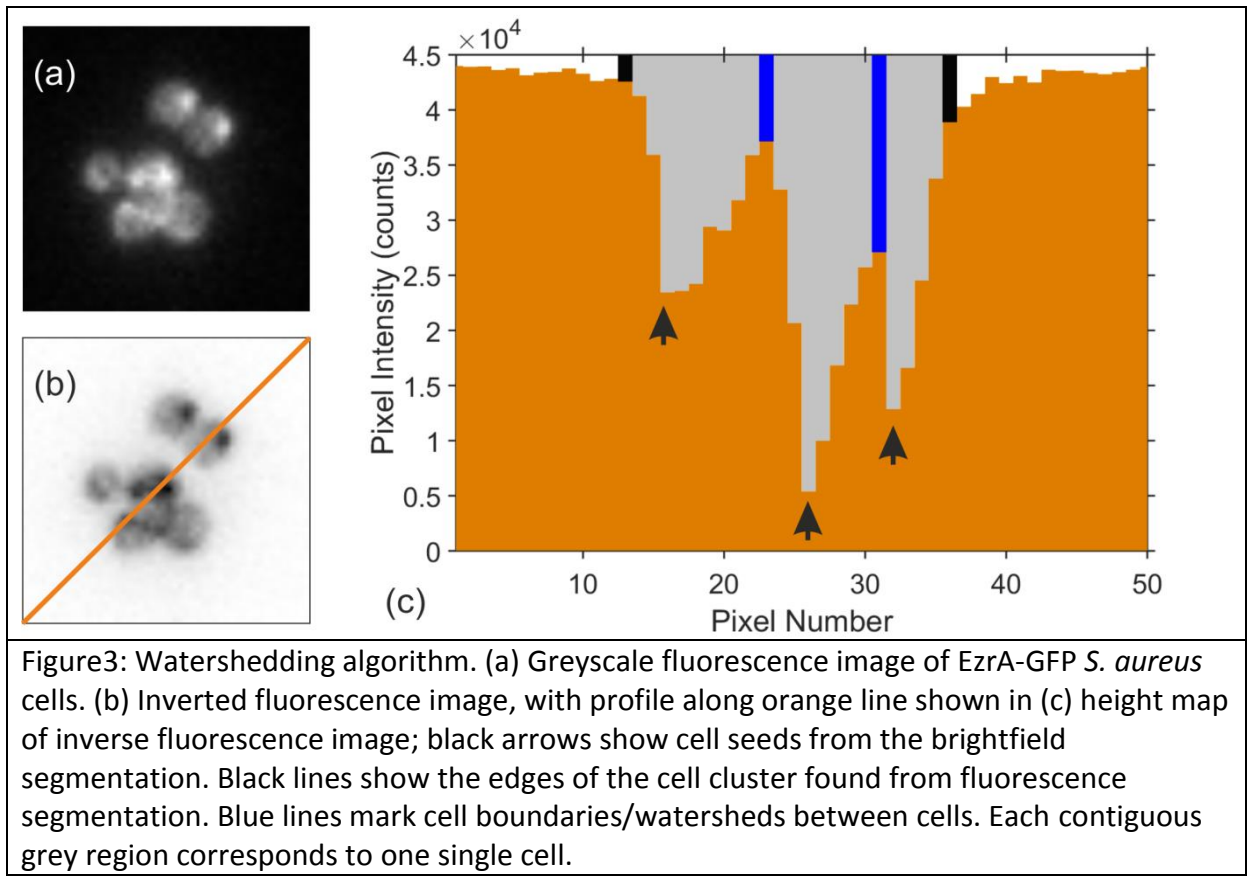

Figure3: Watershedding algorithm. (a) Greyscale fluorescence image of EzrA-GFP *S. aureus* cells. (b) Inverted fluorescence image, with profile along orange line shown in (c) height map of inverse fluorescence image; black arrows show cell seeds from the brightfield segmentation. Black lines show the edges of the cell cluster found from fluorescence segmentation. Blue lines mark cell boundaries/watersheds between cells. Each contiguous grey region corresponds to one single cell.

The pixel positions of each cell centre determined from simple brightfield segmentation are, in the case of *S. aureus* cell clusters, good estimates for the minima of the valleys, the seeds, but several alternative automated methods could in principle also be used to determine their locations (62). The watersheds can now be found by progressively flooding the landscape until the cell regions defined by the seeds merge (62,63). Each seed pixels' eight neighbouring pixels are then sorted from lowest pixel intensity value (i.e. pixels which are most similar to the seed, or, in the analogy of the river basin, closest to the bottom of the valley) to highest. Pixels are considered in turn by looking at which of their eight neighbours have already been assigned to a cell. If a pixel's only labelled neighbours have all been assigned to the same cell it is also assigned to that cell, and its unassigned neighbours are added to the queue at their appropriate heights. If a pixel has two neighbours with different cell assignments, it is considered to lie on the boundary between them, and is therefore defined as part of the watershed. This process is repeated until all pixels in the region have been assigned uniquely to one cell, and a separate 2D ellipse function fit is then applied solely to the pixels within each watershed-defined cellular region.

Imaging analysis framework - 2.Thresholding inside cells (determining EzrA ring localization)

Once the pixels corresponding to single cell foreground images have been identified, as above, it is relatively easy to threshold again inside these segmented cell images. Here, we used Otsu's method (64). Otsu's method is a robust, standard approach which aims to separate a general distribution of values of a parameter into a number of classes through the process of minimising the intra-class variance. In our case here the parameter is that of pixel intensity, and we assume in the simplest hypothesis for there being just two classes, one which corresponds to putative EzrA rings and is manifest as a higher mean pixel intensity due to distinct fluorescently labelled EzrA rings tightly packed at the cell mid-plane, and a second class which comprises more diffusive components of lower mean pixel intensity which corresponds to a combination of background autofluorescence and rapidly diffusing EzrA subunits prior to association with a ring structure. To threshold an image ideally there would be two well separated peaks on the pixel intensity distribution. However, in reality, especially in the case of low signal-to-noise regimes of millisecond Slimfield microscopy, the valley between the two peaks is typically not clearly defined, due to imaging noise and differences in foreground and background pixel distributions. Otzu's method performs well under these conditions, and also offers advantages over other methods such as fitting Gaussian functions (65) or valley sharpening (66) as the peaks are rarely symmetrical Gaussian shapes, and valley sharpening only considers a highly localized area of the distribution, rather than all of the data in an image. GFP labelled EzrA rings appear as relatively brighter objects on a darker cell body background, and so are well suited to Otsu's method with just a single threshold.

Imaging analysis framework - 3. Simulating brightfield and fluorescence images

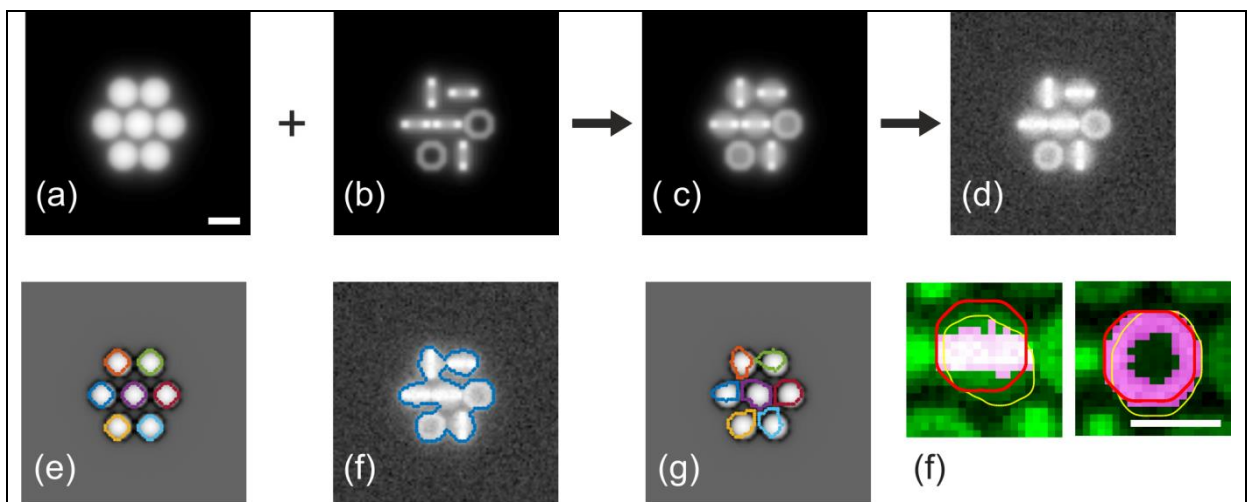

Figure 4: Simulating images (a) uniform cell fluorescence, (b) EzrA ring fluorescence, (c) total fluorescence, (d) noisy fluorescence, (e) segmented brightfield output, here intentionally made to be high contrast to show the distinct cell boundaries, (f) simple segmentation of noisy simulated fluorescence cell images (greyscale) based on a single pixel intensity threshold value (blue), (g) watershed segmentation of simulated noisy fluorescence cell images (coloured lined) using only the brightfield images of cells as seeds (greyscale data), (h) zoom-in of noisy simulated fluorescence image from a single cell (green) with EzrA ring (purple/white), showing an example of a perpendicularly oriented (left panel ) and in-plane (right panel) ring, with simulated segmentation (red) and detected segmentation (yellow) shown. Scale bar 1 µm.

To validate our approach for segmenting the outer cellular boundaries and subcellular morphological features, exemplified by GFP-labelled EzrA rings, we simulated realistic brightfield and fluorescence images of *S. aureus* cells. These included image features of distinct EzrA-GFP rings, a subcellular diffusive background of EzrA-GFP, and a background not associated with GFP which comprised autofluorescence plus camera readout background. Cell background and EzrA-GFP ring fluorescence were modelled by adapting a previously reported method (57) for integrating a model point spread function over a 0.8 µm diameter sphere (fig. 4a) and a randomly orientated, parallel or perpendicular ring (fig. 4b) of diffraction limited width (here set at 0.3 µm) respectively. Here, we retained the same basic tightly-packed pattern and relative orientations of 7 cells in a cluster throughout. These images were summed (fig. 4c) and then scaled to realistic pixel intensity values before realistic levels of pixel noise, trained on experimental fluorescence image data, were added (fig. 4d). Brightfield images (fig. 4e) were simulated by subtracting 0.8 µm diameter rings and circles from each other to generate bright central regions surrounded by dark rings, and were added to a uniform bright background. Images were then segmented using precisely the same algorithms and same parameter set as for real experimental image data (fig. 4f,g,h).

**Results**

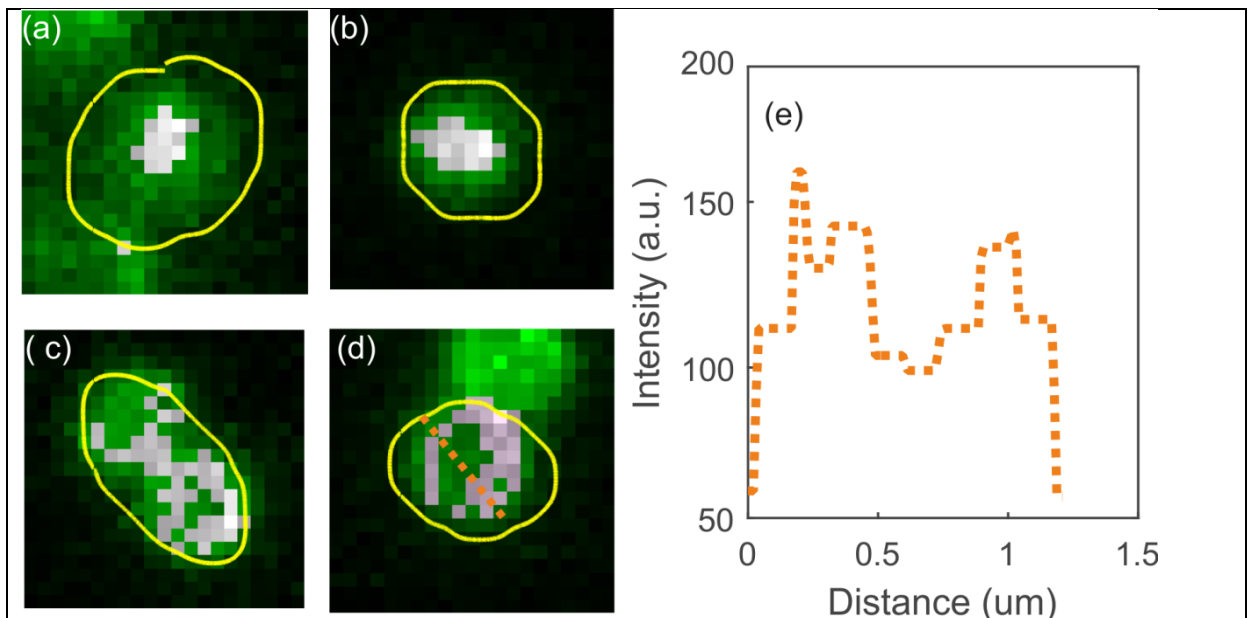

Figure 5: Segmented cells (green), identified putative EzrA-GFP rings highlighted in white. Images show the determined cell boundary (yellow) and the greyscale pixels indicate the pixels associated with putative EzrA-GFP determined from our image analysis framework. (a) & (b): examples of the algorithm detecting division planes in cells. (c) At putative late stages of division the dividing cells have not completely separated and the image segmentation algorithm may categorise these as a single elongated elliptical foreground object. (d) shows a putative EzrA-GFP ring consistent with an orientation of the edge of the ring projecting towards the plane of the camera . (e) shows the pixel intensity profile through the dotted line in (d), indicating a peak-to-peak diameter of ~0.8 µm in this case.

Segmenting cells

Candidate automatically detected cell masks were accepted for subsequent image analysis if the summed pixel area was in the range 0.03-3.00 µm$^2$; this is the equivalent area of a circle whose diameter is in the range 0.2-2 µm, which tallies with prior structural observations of the length scale of *S. aureus* cells during their complete cell cycle. In our proof-of-concept study here this resulted in accepting ~60% of initially detected candidate foreground objects (here, 34 out of 60 initially detected foreground objects from 20 separate fields of view). Example cell boundaries found are shown in fig. 5. Most cell boundaries are slightly elliptical (fig. 5a,b,d) with aspect ratios (ratio of major and minor axis length) close to 1, but a minority had extended boundaries detected with larger aspect ratios closer to 2 (fig. 5c). These examples with extended boundaries are consistent with the appearance of pairs of dividing cells which have been erroneously segmented together as a single cell.

The mean cell length we measure to be 1.2 ± 0.3 µm (± s.d.) (fig. 6a), in good agreement to within experimental error with estimates (67,68), though as noted from super-resolution studies there can be significant variation of cell length depending on the specific stage in the cell cycle (24). The majority of our data have a major axis which is 30-50% longer than the minor access, indicating a mean cell aspect ratio of 1.4 ± 0.3 (fig. 6b,c). The recent super-resolution investigations of Monteiro *et al.* (24) made measurements of the aspect ratio and cell dimensions using structured illumination microscopy images of vancomycin-labelled peptidoglycan in *S. aureus*. Here they measured similar ranges of aspect ratio to within experimental error (Student *t* test, P<0.001) for cells in the P2 and P3 phases when the cells were dividing and EzrA was located at the division plane.

We used simulations to determine the cell boundary determination error of our image analysis framework, defined as the mean deviation of the cell boundary from the watershedded boundary. The deviation for any point on the detected boundary is defined as the absolute value of the shortest distance between the simulated and detected boundary. Figure 6d shows the distribution of boundary errors for 70 simulated cells with different levels of simulated noise. Experimentally measured pixel intensity values for cells had a standard deviation noise of as a much as ~50% of the mean cell background pixel intensity, but with a more typical level of ~20% (Supplementary Information). In our simulations we found that although brightness variation does generate more outliers the boundary determination was relatively insensitive to these relatively large fluctuations in pixel intensity, culminating in a typical boundary precision in the range 100 - 300 nm.

In our simulations, 100% of cells were detected successfully. However, in real experimental data the same image analysis framework rejected up to ~40% of initially detected candidate foreground objects. Failure to detect cells may occur in principle when insufficient fluorescence signal is present, or if there is some misalignment between the brightfield and fluorescence images, or if the measured cell mask area is beyond the imposed area acceptance range limit. We observed examples of all three categories in our data. Gene expression both for the EzrA protein and for natively autofluorescent proteins is stochastic in nature, and so there will inevitably be a minority of cells which have low intrinsic levels of fluorescence too close to the level of camera readout noise to permit robust image segmentation. Similarly, misalignment between fluorescence and brightfield images more commonly occurs when using differential interference contrast (DIC), since a Wollaston prism slider is placed just under the objective lens and can often result in mechanical based misalignment of the sample (e.g. slight knocks on the sample stage) in addition to the polarization optics resulting in a lateral shift of the image on the camera detector. Although DIC was not used here, we included some accidentally misaligned data intentionally (as revealed upon close inspection of fig. 2 for example) to demonstrate that our image analysis framework is in general sufficiently robust to cope with minor misalignment issues.

The most relevant rejection category we found for our data was on the basis of the area acceptance range limit. Here, we set the upper limit to correspond to an effective cell diameter of 2 µm to therefore exclude the majority of clusters of >1 cell which our algorithm had failed to segment into individual cells. The majority of rejected cell masks were of this type. However, some were rejected by being less than the lower area limit threshold, equivalent to an effective cell diameter of 0.2 µm. These included images which were consistent with being cell fragments from dead cells, however there were also a minority of instances in which the primary segmentation step would detect the outline of the EzrA ring itself as opposed to the outline of the cell boundary – these were instances of a minority of cells which had a much lower intrinsic level of fluorescence for the cell background. These detection limits do not preclude using our image analysis framework for determination of cell lineages, since the same 60% acceptance level will be propagated over time. And it should be noted that our aim here was not to achieve 100% detection efficiency, but rather to intentionally have a stricter acceptance policy to increase the confidence for data interpretation of the accepted segmented cell images.

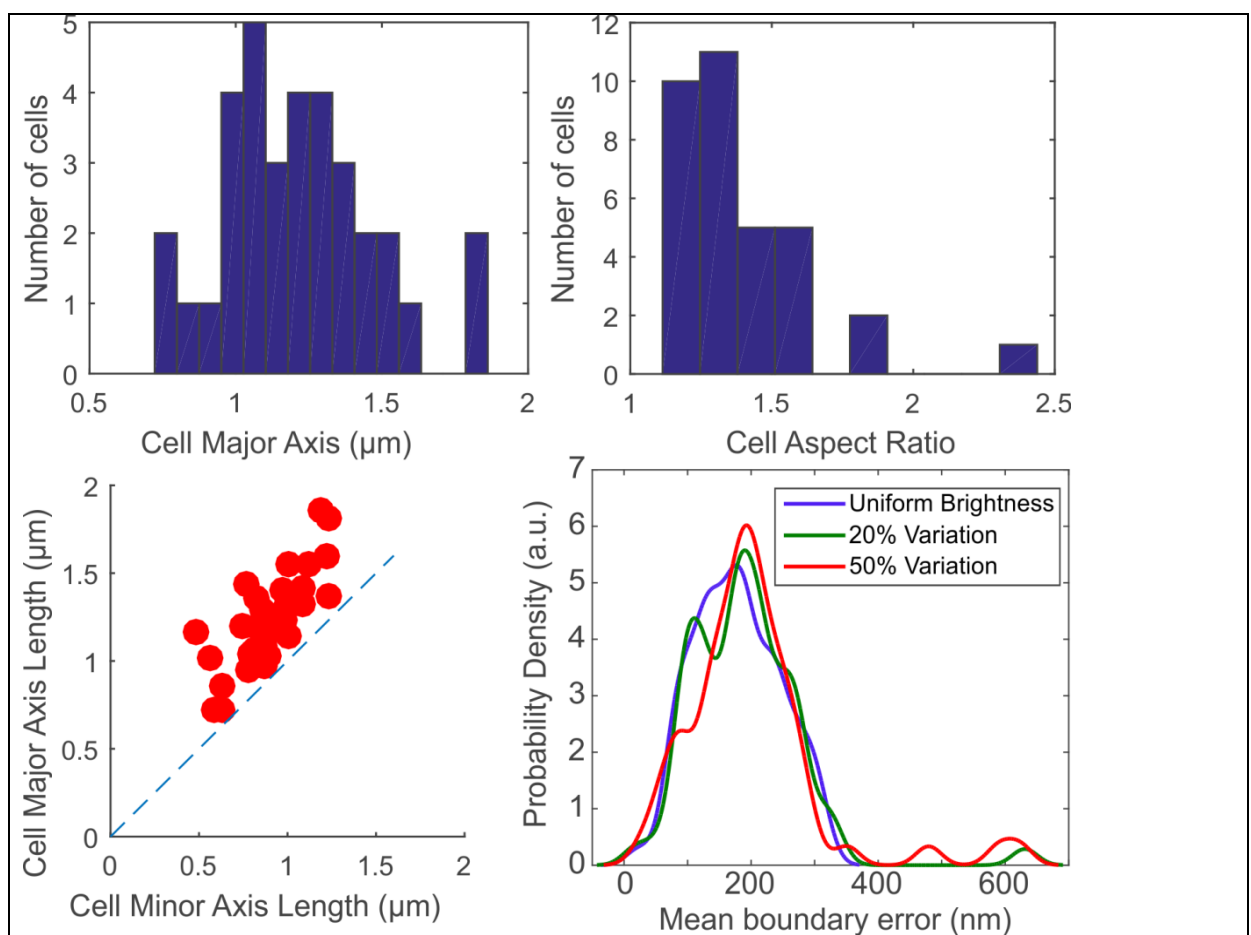

Figure 6: Top Left: Distribution of cell major axis lengths, Top Right: aspect ratio, Bottom Left scatter plot of cell major axis length against minor axis length. The dotted line indicates an aspect ratio of exactly 1 (i.e. a circle), showing that the majority of cells are elongated. Bottom Right: distribution of mean boundary error for different variations in cell brightness, compiled from 70 different simulated cell images of mixed in-plane and perpendicular orientations using either a uniform level of background noise with 0% fluctuation (blue), or random Gaussian background noise using a standard deviation value of 20% (green) or 50% (red) of the mean background intensity level.

Identifying EzrA rings

A range of different shaped regions of fluorescently-labelled EzrA can be found by applying pixel thresholding inside the cellular boundary regions (fig. 3). Elliptical fits to these regions produce some thin extended ellipses but also more circular fits. The distribution of aspect ratios of these pixel regions and a scatter plot of major against minor axis length (fig. 7a,b) show that ~50% of cells have extended structures with aspect ratios far in excess of 1, almost as high as 4, consistent with EzrA rings perpendicularly oriented to the image plane. We confirmed this by using simulations of perpendicularly oriented and in-plane rings. Unlabelled (wild type) cells, do not contain any visible rings in fluorescence image (Supplementary Fig. 1). The line profile through a putative in-plane ring shows a clear double peak (fig. 5e), as would be expected. The remaining structures are more circular, either corresponding to in-plane rings or a completely delocalised diffusive EzrA-GFP. These can be distinguished on the basis of their estimated areas as a function of major axis length, which accounts for cell orientation projection effects onto the camera detector. Figure 7c summarises how the area, *A*, varies as a function of major axis length, *2r*, for a fixed ring width, *w*, for a continuous circular region as produced by delocalised diffusive EzrA-GFP molecules, an in-plane ring and an ellipse produced by a perpendicularly oriented EzrA-GFP ring. Most of the accepted cell foreground objects are consistent with the presence of a ring, although a few are consistent with continuous but truncated localised EzrA-GFP structures, indicative more of short protofilaments than rings, and suggesting that these cells are not actively dividing.

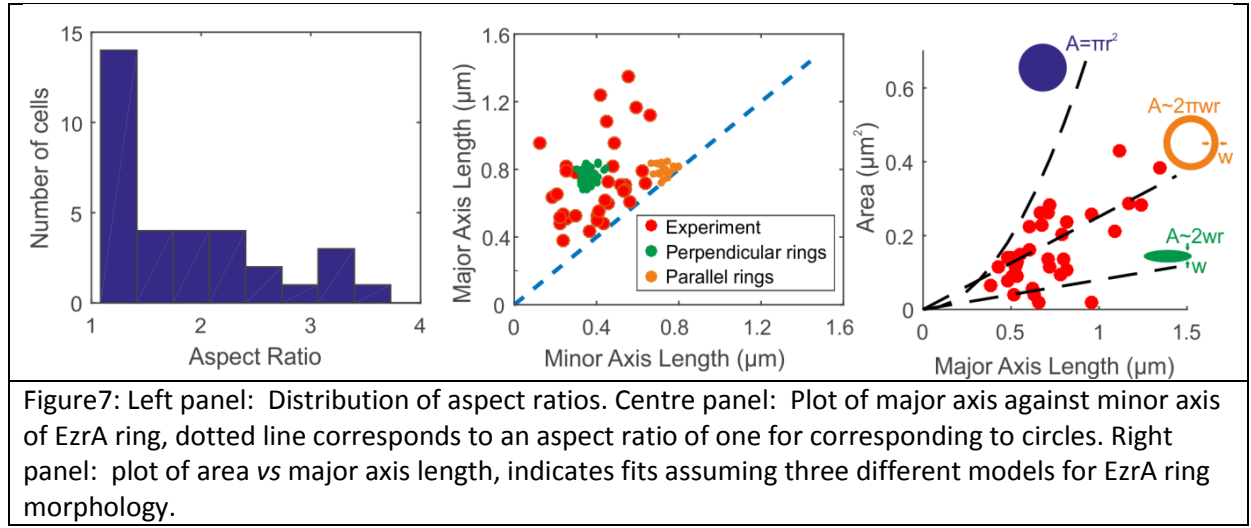

Figure7: Left panel: Distribution of aspect ratios. Centre panel: Plot of major axis against minor axis of EzrA ring, dotted line corresponds to an aspect ratio of one for corresponding to circles. Right panel: plot of area *vs* major axis length, indicates fits assuming three different models for EzrA ring morphology.

**Discussion**

Our straightforward image analysis framework detects cells and characterises their size and shape assuming an ellipsoidal model for the general 3D shape of *S. aureus* cells, manifest as a 2D ellipse on an image projection. It then detects bright pixels inside the cell corresponding to EzrA rings, characterises their shape and determines their orthogonality to the image plane. Our method is valuable for investigating *S. aureus* cells, which do not move apart following the conclusion of the cell division process. The analysis framework can be extended to study other fluorescently-labelled proteins in *S. aureus*, but also in other clustering cells since it does not require the foreground objects to be spatially separated by regions of background pixels. The watershedding method, which here uses brightfield cell centres as seeds, is robust to imaging data for which the brightfield image is not precisely aligned with the fluorescence image. Here, we are not claiming to have developed any single novel image segmentation method *per se* - our image analysis framework here uses existing, standard methods, quite clearly. Rather, we use these in combination to create a framework which has previously never been applied to challenging data from millisecond images of live cells which have morphologically heterogonous subcellular features, as exemplified by the pathogen *S. aureus* with subcellular EzrA ring structures.

The aspect ratios we find for cells are in agreement with those found by Monteiro *et al.* (24), indicating that super-resolution imaging is not necessarily required to extract this parameter. Using the autofluorescence of the cell potentially leaves other spectrally delimited channels open for protein studies (i.e. multi-colour fluorescence imaging). We find

EzrA rings are localised to the division plane in agreement with the expected distributions during cell division.

Other studies have required manual segmentation (53) or relied on super-resolution images (24) to achieve similar results. Our simple framework is fully automated and does not require costly and potentially damaging super-resolution imaging. However it is still compatible with super-resolution microscopy images, but also with millisecond microscopy such as Slimfield illumination as well as other time-resolved fluorescence localization microscopy tools which enable tracking of single-molecule complexes (27,56,69), for example to enable quantification of protein copy number in Erza rings using convolution modelling (57).

**Conclusion**

We have constructed a simple bespoke automated image analysis framework using a combination of several standard approaches which enables segmentation of individual *S. aureus* live cell images within cell clusters, and can detect the cell division planes using fluorescently-labelled EzrA protein as a marker, from millisecond sampled images. The framework can be used to investigate cell aspect ratios, other labelled proteins that may be involved in division in *S. aureus*, and it may also have wider applicability for studying other clustering cells since it does not require cells to be separated by non-cellular background pixels. *S. aureus* is an increasing healthcare problem, particularly methicillin resistant and vancomycin resistant strains. It thus has value towards gaining new insight into the operating mechanism of cell division to facilitate the development of future new cell division targeting antibiotics.

**Acknowledgements**

We acknowledge technical assistance for cell preparation (Robert Turner) and image acquisition (Richard Nudd). The research was supported by the White Rose Consortium (WRC), the Biological Physical Sciences Institute (BPSI), University of York and the Medical Research Council (grant number MR/K01580X/1).